\documentclass[amsmath,amssymb,prc,final,showpacs,twocolumn]{revtex4-1}

\usepackage{bm,hyphenat,xspace}
\usepackage{graphicx,epsfig}
\usepackage{color}

\newcommand {\mcj}{\mathcal{J}}

\renewcommand{\Re}{\mathrm{Re}}
\renewcommand{\Im}{\mathrm{Im}}

\begin{document}


\title {Tetraneutron resonance in the presence of a dineutron}

\author{A.~Deltuva}
\email{arnoldas.deltuva@tfai.vu.lt}
\affiliation
{Institute of Theoretical Physics and Astronomy,
Vilnius University, Saul\.etekio al. 3, LT-10257 Vilnius, Lithuania
}
\author{R. Lazauskas}
\email{rimantas.lazauskas@iphc.cnrs.fr}
\affiliation{IPHC, IN2P3-CNRS/Universite Louis Pasteur BP 28, F-67037 Strasbourg Cedex 2, France}


\received{\today}
\begin{abstract}
\begin{description}
\item[Background] Several previous studies provided contradicting
results for the four-neutron system, some claiming the existence
of a $0^+$ near-threshold resonance, others denying presence of
any observable resonant states.
\item[Purpose] Since most of the
studies employed enhanced  two-neutron interactions to follow the
evolution of an artificially bound state into a continuum one, we
examine several enhancement schemes that produce a bound dineutron
as well.
 \item[Methods] We
study the four-neutron system by solving exact four-particle
equations. By varying the interaction enhancement factor we
calculate  two-dineutron scattering phase shifts and cross
sections.

\item[Results] When the same enhancement factor is used in all
partial waves, a bound tetraneutron emerges together with a
strongly bound dineutron. Furthermore, such a $0^+$ tetraneutron
evolves not into a resonance but into a virtual state. Weak
enhancement of $S$ waves together with strongly enhanced higher
waves is needed for the emergence of the resonant state. Anyhow
the resonant behavior disappears well before reaching the physical
interaction strength. \item[Conclusions] The interaction
enhancement scheme using the same factor for all waves, employed
in several previous works, is misleading for the search of $0^+$
resonance as only a virtual state can emerge. Evolution of a bound
tetraneutron into a resonance via an intermediate virtual state is
possible with strong enhancement of higher two-neutron waves.
\end{description}
\end{abstract}

 \maketitle

\section{Introduction}

A possible experimental observation of the tetraneutron resonance
\cite{PhysRevLett.116.052501} triggered a number of theoretical
studies
\cite{PhysRevLett.117.182502,PhysRevLett.118.232501,PhysRevLett.119.032501,PhysRevC.93.044004,deltuva:18b}.
Their results are highly contradictory: a resonance with total
angular momentum and parity $\mcj^\Pi = 0^+$ just 1 - 2 MeV above
threshold was obtained using a
 harmonic oscillator representation of the continuum
\cite{PhysRevLett.117.182502} and the bound-state quantum Monte
Carlo with extrapolation to the continuum
\cite{PhysRevLett.118.232501}. Calculations based on the no-core Gamow
shell model \cite{PhysRevLett.119.032501} and the density matrix
renormalization group \cite{PhysRevLett.119.032501} approaches
have not found any narrow resonant states and tolerate only the 
presence of  broad structures with a width of at least 4 MeV,
while their positions could not be determined accurately. On the
other hand, rigorous solution of four-particle Faddeev-Yakubovsky
(FY) equations in the coordinate- \cite{PhysRevC.93.044004} and
momentum-space \cite{deltuva:18b} representations predict no
observable tetraneutron resonance, consistently with the
pioneering studies on this subject
\cite{sofianos_4n,lazauskas:4n}. The aim of the present work is to
shed some light on the origin of these disagreements.

We do not discuss the details of our four-particle bound-state  and scattering
calculations that can be found in
 Refs.~\cite{PhysRevC.93.044004,lazauskas:4n,deltuva:18b}
and in references therein. In Sec.~\ref{s2} we present the study of the
artificial four-particle system using two different force schemes.
The summary and conclusions are contained in  Sec.~\ref{s3}.

\section{Results \label{s2}}

Apart from the work~\cite{PhysRevLett.117.182502}, none of the
aforementioned theoretical studies were able to identify
tetraneutron resonances by performing direct calculations
considering the physical values of the nuclear interaction. In the
studies~\cite{sofianos_4n,lazauskas:4n,PhysRevLett.119.032501,PhysRevC.93.044004,deltuva:18b}
no tetraneutron complex-energy states were observed sufficiently
close to the physical energy axis, whereas Monte-Carlo techniques
used in \cite{PhysRevLett.118.232501} are unsuited to determine
the positions of unbound states directly. All of the
aforementioned studies \footnote{In
Ref.~\cite{PhysRevLett.117.182502} this was an auxiliary method}
therefore tried to enhance the nuclear interaction to make the
four-neutron ($4n$) system bound, and then follow its evolution to
the physical strength of the interaction.

However, important differences were present in the details:
 Refs.~\cite{PhysRevLett.117.182502,PhysRevLett.118.232501,PhysRevLett.119.032501}
 enhanced the neutron-neutron ($nn$) potential in all partial waves
by the same factor $f$, Refs.\cite{PhysRevC.93.044004,lazauskas:4n}
added an attractive $3n$ or $4n$ force, Ref.~\cite{deltuva:18b}
enhanced the $nn$ potential in partial waves with orbital momentum
$L>0$ by the same factor $f_L$
 but kept the physical strength $f_0=1$ for the ${}^1S_0$ force.
The strategies of
Refs.~\cite{PhysRevC.93.044004,lazauskas:4n,deltuva:18b} had a
particular goal to avoid binding dineutrons. In contrast, when the
$nn$ force is enhanced also in the ${}^1S_0$ wave as in
Refs.~\cite{PhysRevLett.117.182502,PhysRevLett.118.232501,PhysRevLett.119.032501},
very soon a bound dineutron is generated. The presence of a bound
dineutron with energy $E_d < 0$ relative to the free-particle
threshold has several consequences for the $4n$ system. First, the
lowest scattering threshold is no longer at the energy
$E_{4n}=0$ but at the two-dineutron threshold $E_{4n}= 2E_d < 0$.
Thus, negative energy  $4n$ states  at $ 2E_d < E_{4n} < 0$ are
not bound but continuum states that may be realized in the
scattering of two dineutrons, only those with $ E_{4n} < 2 E_d$
are bound states. Second, the presence of an additional threshold
changes the structure of the complex energy plane of  $4n$ states,
further complicating trajectories of $4n$ states.

To reveal shortcomings of the approach of enhancing the ${}^1S_0$
$nn$ force, we study the $4n$ system using rigorous FY equations
in the coordinate-space \cite{PhysRevC.93.044004} and FY
equivalent momentum-space Alt, Grassberger, and Sandhas (AGS)
equations for transition operators \cite{deltuva:18b}. We start
with the model enhancing all partial waves by the same factor,
i.e., $f_0 = f_L = f$. As it is well known, a bound dineutron
emerges in the ${}^1S_0$ partial wave at $f_0 \approx 1.1$; this
critical value of $f_0$ slightly varies with the $nn$ potential,
see Ref.~\cite{lazauskas:3n} for a number of realistic potentials.
The energy of the dineutron $E_d$ relative to the free neutron
threshold rapidly decreases with increasing $f_0$, i.e., it
becomes bound more tightly. To support a bound tetraneutron, i.e.,
the one with $ E_{4n} < 2 E_d$, significantly larger enhancement
is needed; it also depends on the potential. A bound tetraneutron
emerges first in the $\mcj^\Pi = 0^+$ state at $f\approx 2.4 $ for
several local realistic potentials such as Nijmegen II or Argonne
V18, where the dineutron energy is below $-20$ MeV
\cite{lazauskas:phd}. In order to make comparison with
Refs.~\cite{PhysRevLett.117.182502,PhysRevLett.118.232501,PhysRevLett.119.032501}
that used  soft nonlocal potentials, our following results will be
based on the next-to-leading order (NLO) chiral potential
\cite{PhysRevLett.115.122301} as in
Ref.~\cite{PhysRevLett.118.232501}. With this model and the space
of  Ref.~\cite{deltuva:18b} the tetraneutron becomes bound at
$f=2.665$ where $E_d = -12.01$ MeV and $ E_{4n} = -24.02$ MeV. At
lower $f$ values the $4n$ system cannot be bound, the
lowest-energy $4n$ state is the two-dineutron state. However,
Refs.~\cite{PhysRevLett.118.232501,PhysRevLett.119.032501} report
a bound tetraneutron at threshold $ E_{4n} \to -0$ with $f$ being
as low as 1.3 \cite{PhysRevLett.118.232501} or 1.6
\cite{PhysRevLett.119.032501}. Those $4n$ states in
Refs.~\cite{PhysRevLett.118.232501,PhysRevLett.119.032501} have
zero width, which is not compatible with the fact that they are
embedded in the two-dineutron continuum. Nevertheless, one might
still expect that those are resonant states that  evolved from the
bound state with decreasing $f$, and thereby could provide
estimation for the real part of the $4n$ resonance energy.
 To verify this conjecture we study the $f$-evolution of two-dineutron
scattering states using rigorous treatment of the four-particle continuum
as provided by AGS equations.

\begin{figure}[!]
\begin{center}
\includegraphics[scale=0.68]{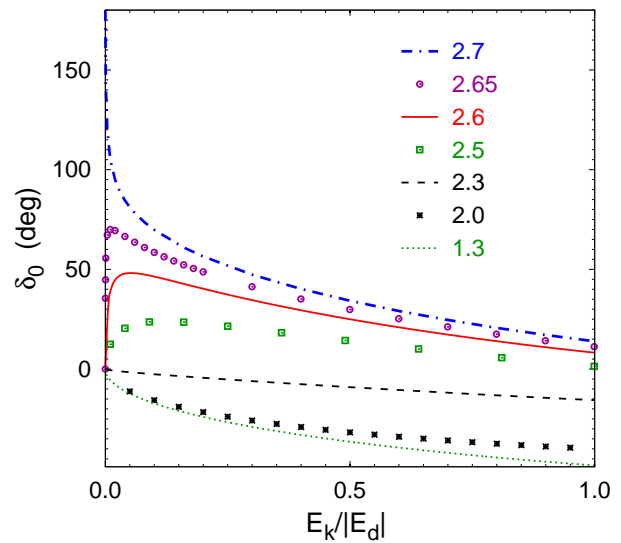}
\end{center}
\caption{\label{fig:ff}
$J^\Pi = 0^+$ phase shift for the
scattering of two artificially bound dineutrons as a function
of the center-of-mass kinetic energy $E_k$.
NLO potential enhanced by a factor $f$  as
indicated in the plot was used.}
\end{figure}

\begin{figure}[!]
\begin{center}
\includegraphics[scale=0.68]{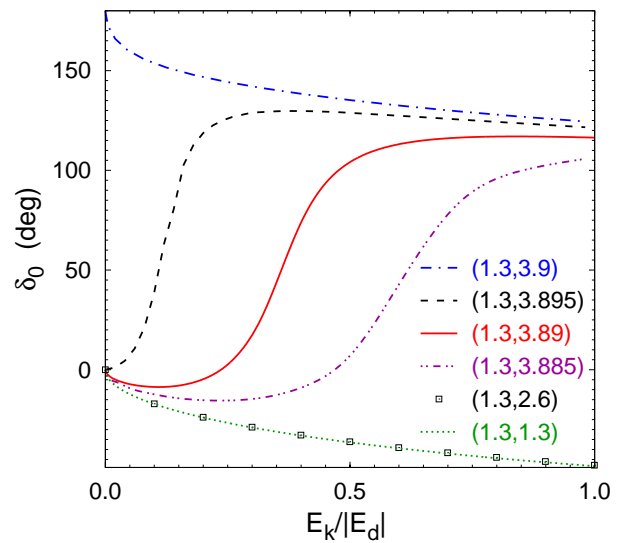}
\end{center}
\caption{\label{fig:fl}
$J^\Pi = 0^+$ phase shift for the
scattering of two artificially bound dineutrons as a function
of the center-of-mass kinetic energy $E_k$.
Enhanced NLO potential
with $S$- and higher-wave factors  $(f_0,f_L)$ as
indicated in the plot was used.}
\end{figure}

In Fig.~\ref{fig:ff} we show dineutron-dineutron $0^+$ phase shift
$\delta_0$ as a function of the kinetic energy $E_k$ in the
center-of-mass (c.m.) frame at $f=2.7$, 2.65, 2.6, 2.5, 2.3, 2.0,
and 1.3. There is a clear qualitative difference between the
$f=2.7$ and 2.65 phase shift results at low energies: the former
monotonically decreases from 180 deg at threshold whereas the
latter rapidly increases from 0 deg at threshold exhibiting a bump
of 70 deg. This signals the emergence of a bound $4n$ state at $f$
between 2.65 and 2.7, fully consistent with the result of 2.665
from the direct bound state calculation. However, the most
important observation is that at $f=2.65$, 2.6, and 2.5 where one
could expect the bound state to evolve into a resonance, no
resonant phase shift behavior is observed. Note that qualitatively
the same behavior is seen in the realistic  ${}^1S_0$ two-nucleon
phase shift, reflecting the presence of   the well-known virtual state
near threshold. Indeed,  at   $f=2.65$, 2.6, and 2.5 the
two-dineutron phase shift energy dependence is  consistent with
the presence of a virtual $4n$ state that with decreasing $f$
 moves away from the  two-dineutron threshold into the unphysical sheet
of the complex energy plane.
Reducing $f$ further,  the virtual state
becomes too far from the threshold to have a visible effect,
 while the phase shifts   approach the universal ones
obtained for two fermionic dimers in the unitary limit \cite{deltuva:17d},
signaling an effective repulsion
between the two dineutrons. For example, at $f=1.3$ the deviation
from the universal results \cite{deltuva:17d} is below 5\%.
Thus, in the $nn$ force enhancement scheme
 $f_0 = f_L = f$ the $0^+$ tetraneutron bound state evolves not into a resonance
but into a virtual state, rendering this scheme entirely misleading
for the $0^+$ resonance study.

The evolution of a bound into a virtual state is typical for
$S$-wave two-body systems \cite{res_cpl} unless there is
a sufficiently high potential barrier leading to the appearance of
a resonance, as shown in the example of Ref.~\cite{PhysRevLett.118.232501}
based on a two-Gaussian potential. The angular-momentum barrier is
always present in $L>0$ waves, resulting in the evolution of a bound state
into a resonance. In the $4n$ systems both $L=0$ and $L>0$ waves are present,
but their balance at $f_0 = f_L$ clearly favors a virtual state,
 not a resonance.
The $S$-wave dominance in the physical $f=1$ case is demonstrated
also in Ref.~\cite{deltuva:18b}. 

In the following we will show a
counterexample where the $0^+$ tetraneutron bound state evolves
into a resonance seen in the two-dineutron scattering. To achieve
this goal one needs to break the $S$-wave dominance, using larger
enhancement factor $f_L$ for $L>0$ waves as compared to $f_0$. In
our example we take $f_0=1.3$. In that case the dineutron energy
is $E_d = -0.316$ MeV. To support a bound tetraneutron with
$E_{4n} < 2 E_d$,
 significantly stronger enhancement $f_L =3.898$ is needed for higher waves.
Results for two-dineutron phase shifts obtained around these
values of $(f_0,f_L)$  are presented in Fig.~\ref{fig:fl}. There
is clear qualitative difference between $f_L = 3.9$ and 3.895
results, signaling the presence of a bound state in the former
case and a resonance in the latter case, consistently with direct
bound-state calculations.  With decreasing $f_L$ the resonance
rapidly moves to higher energy and becomes broader, as  is
evident from the energy dependence of $f_L =3.895$, 3.89, and
3.885 two-dineutron phase shifts shown in Fig.~\ref{fig:fl}. Thus,
under these conditions of $f_L$ being considerably larger than
$f_0$, the  $0^+$ tetraneutron bound state evolves into a resonance
as there is a strong contribution of higher partial waves creating
an effective repulsive barrier via centrifugal terms. However,
with decreasing $f_L$ the dominance of $S$-waves is restored well
before reaching the $f_L = f_0$ such that the tetraneutron pole on
the unphysical sheet moves deeply below the two-dineutron
threshold, without having a visible effect on the scattering
processes, as already known from $f=1.3$ results  in Fig.
~\ref{fig:ff}. In fact,
 already at $f_L= 2f_0 = 2.6$ the phase shift results are very close
to those of  $f_L= f_0 = 1.3$ and thereby also to the universal ones.

\begin{figure}[!]
\begin{center}
\includegraphics[scale=0.68]{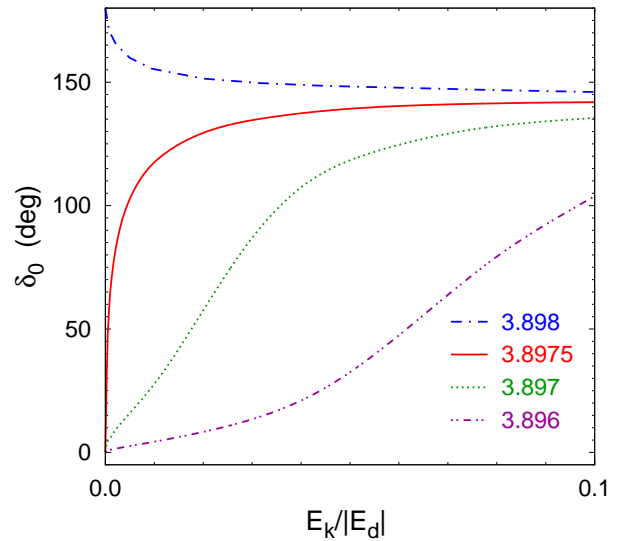}
\end{center}
\caption{\label{fig:crit}
$J^\Pi = 0^+$ phase shift for the
scattering of two artificially bound dineutrons as a function
of the center-of-mass kinetic energy $E_k$.
Enhanced NLO potential
with $f_0=1.3$  and different higher-wave factors  $f_L$
indicated in the plot was used.}
\end{figure}

\begin{figure}[!]
\begin{center}
\includegraphics[scale=0.68]{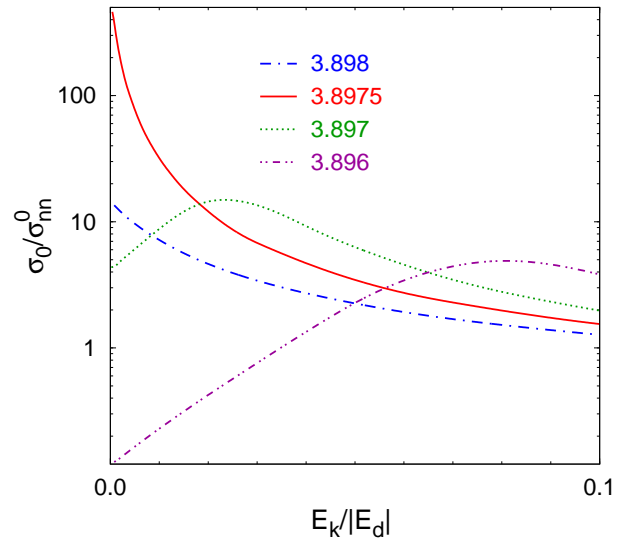}
\end{center}
\caption{\label{fig:cs}
$J^\Pi = 0^+$ total cross section $\sigma_0$ for the
scattering of two artificially bound dineutrons as a function
of the center-of-mass kinetic energy $E_k$.
Enhanced NLO potential
with $f_0=1.3$  and different higher-wave factors  $f_L$
indicated in the plot was used.}
\end{figure}

The regime very close to the critical point  where the
tetraneutron becomes unbound deserves special consideration. In
Fig.~\ref{fig:crit} we show two-dineutron phase shifts $\delta_0$
obtained with $f_0=1.3$ and $f_L=3.898$, 3.8975, 3.897, and 3.896
in a narrow energy regime $E_k < |E_d|/10$ very close to the
threshold and in Fig.~\ref{fig:cs} the corresponding $0^+$ total
cross sections $\sigma_0$; the latter are normalized by the
two-neutron zero-energy cross section $\sigma^0_{nn}$
calculated with $f_0=1.3$.
 Clearly, $f_L=3.898$ corresponds to a bound state,
while $f_L=3.897$ and 3.896 cases appear to be resonant. However,
at $f_L=3.8975$  the energy-dependence of phase shift and cross
section is  consistent with the presence of a virtual $4n$ state
very near to the threshold, not a bound state or resonance. Thus, in
the case of the strong $L>0$ wave enhancement the bound state
first evolves into a virtual state that, however, further evolves
into a resonance.

The threshold behavior of the cross section
can be read off from the two-dineutron effective range expansion
\begin{equation} \label{eq:efr}
k \cot{\delta_0} = -\frac{1}{a_{dd}} + \frac12 \, r_{dd}k^2 + o(k^4)
\end{equation}
with the on-shell momentum $k$, the scattering length $a_{dd}$,
and the effective range parameter $r_{dd}$.
Namely,
$\sigma_0 \sim 1/|k \cot{\delta_0}-ik|^2$
 is increasing (decreasing) at $E_k=0$ if the ratio
$r_{dd}/a_{dd}$ is above (below) 1. Since in the regime
$f_L < 3.898$ with no bound tetraneutron $a_{dd}$ is negative,
only negative $r_{dd}$ with $|r_{dd}| > |a_{dd}|$ leads to increasing
 $\sigma_0$ as seen for  $f_L \le 3.897$
resonances in Fig.~\ref{fig:cs}.
Though $r_{dd} < 0$  in the whole regime  $f_L \in [3.885,3.9]$,
large $|a_{dd}|$ at $f_L=3.8975$ leads to decreasing $\sigma_0$.
The values are $a_{dd} \approx -29\, a_0$ and  $r_{dd} \approx -9\, a_0$
at $f_L=3.8975$, while
$a_{dd} \approx -2\, a_0$ and  $r_{dd} \approx -14\, a_0$
at $f_L=3.897$,
expressed in terms of the two-neutron scattering length
$a_0 = 12.72$ fm (taken  at $f_0=1.3$).
Thus, in this regime $r_{dd}$ is not only negative, but also
of unnaturally large absolute value. As will be discussed below,
this feature is essential for the appearance of resonant behavior.
Note, that in all studied  $f_L= f_0$ cases $r_{dd}$ is positive,
as in the universal regime \cite{deltuva:17d}.

\begin{figure}[!]
\begin{center}
\includegraphics[scale=0.36]{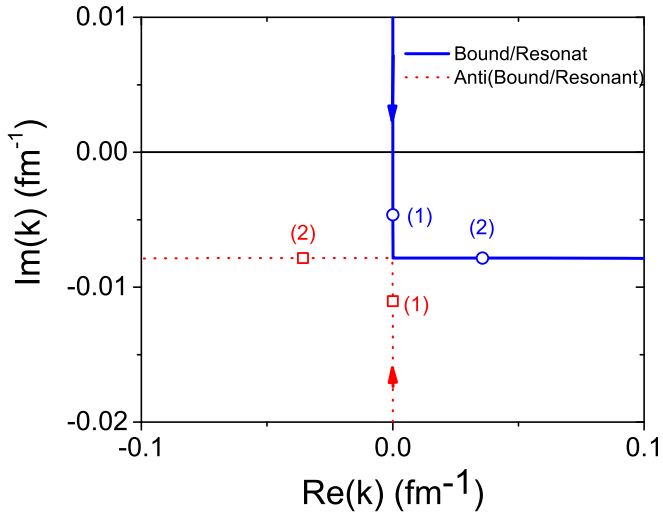}
\end{center}
\caption{\label{fig:trjk} Two-body S-matrix pole trajectories in the
momentum manifold obtained varying the strength $f_0$ of the
two-Gaussian potential in Eq.~(\ref{eq:ff}).
Arrows indicate the direction of reducing $f_0$.
Points (1) and (2) are described in text.}
\end{figure}

\begin{figure}[!]
\begin{center}
\includegraphics[scale=0.36]{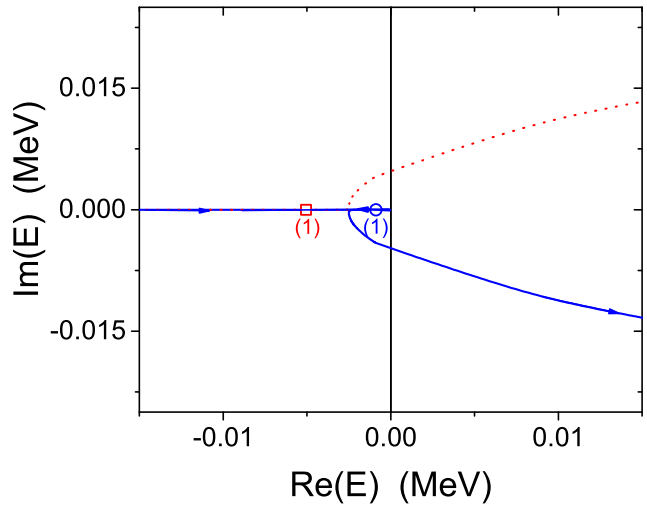}
\end{center}
\caption{\label{fig:trje} Two-body S-matrix pole trajectories in the
energy manifold close to threshold.
Notation is taken from Fig.~\ref{fig:trjk}. Point (2) is outside the plot.
}
\end{figure}

\begin{figure}[!]
\begin{center}
\includegraphics[scale=0.36]{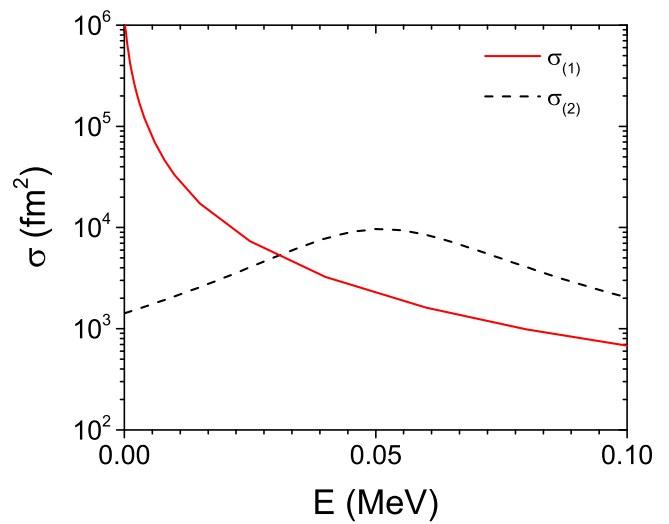}
\end{center}
\caption{\label{fig:cs2} Total cross sections for two-body scattering
calculated using potentials in Eq.~(\ref{eq:ff}). Two cases are
considered, corresponding to the marked points in
Fig.~\ref{fig:trjk}: virtual state (1) and near-threshold resonance
(2).}
\end{figure}

This kind of  bound state evolution into a resonance via the
virtual state may also appear in two-body systems for the $L=0$
($S$-wave) state generated by the short range potential with special
properties. A general feature of the S-matrix pole trajectory,
represented in the momentum $(k)$ manifold, is that it moves down along
the imaginary axis when reducing the attraction. In this way the
bound state with $\Im(k) > 0$ crosses the threshold becoming a
virtual state with $\Im(k) < 0$. Nevertheless, in some particular
cases, quite soon this S-matrix pole may move into the fourth
quadrant of the complex momentum plane: first, as long as $-\Im(k)
> \Re(k)$,
 appearing as a subthreshold resonance
($\Re(E)<0$)
and  then, once $\Re(k)>-\Im(k)$,  turning into a
resonance above the threshold ($\Re(E)>0$). 
Such a non-standard behavior is determined by the presence of a
large negative effective range parameter $r_0$  in the effective
range expansion of the form (\ref{eq:efr}), which can be generated
by employing a potential containing a repulsive barrier. To study
this case, we adapt the two-Gaussian potential from
 Ref.~\cite{PhysRevLett.118.232501}:
\begin{equation}
 V(r)=f_0 \{-1000e^{-(r/0.4981)^2}+865e^{-[(r-0.9972)/0.2877]^2} \}.
 \label{eq:ff}
\end{equation}
Here the potential $V(r)$ is in units of MeV, whereas the distance
between the neutrons $r$ is in fm; neutron mass is fixed by
$\hbar^2/m_n=41.4425$ MeV$\cdot$fm$^2$. We study the evolution of
the S-matrix pole by varying the potential enhancement factor
$f_0$. The S-matrix pole trajectory generated in this way is presented
in Figs.~\ref{fig:trjk} and \ref{fig:trje}, along with the
trajectory of unphysical antibound/antiresonant state. The
physical and unphysical poles collide at $k \approx i/r_0$, where
$r_0$ approaches half of the two-body scattering length $a_0$.
This can be understood from the effective range expansion,  which
is typically a good approximation near the unitary limit $1/a_0 =
0$ and is suitable to estimate the positions of the near-threshold
S-matrix poles given by
\begin{equation} \label{eq:polear}
k = \frac{1}{r_0} \, (i \pm \sqrt{2r_0/a_0 -1}).
\end{equation}
Given a large negative $a_0$ for a near-threshold virtual state,
large  negative $r_0$ ($\sim -125$ fm in the present example)
 ensures that the two poles collide on the negative
imaginary momentum axis in close vicinity of the threshold
 and then scatter by angle $\pi/2$, see
Fig.~\ref{fig:trjk}. In Fig.~\ref{fig:cs2} cross sections are
plotted at two selected trajectory points: at $f_0^{(1)}=1.004325$
generating a virtual state at $k=-0.0046i$ fm$^{-1}$,  and at
$f_0^{(2)}=1.0042$ generating a resonance at $k=0.0357-0.0078i$.
The total cross section, dominated by the presence of the virtual state,
decreases with energy from $E=0$. On the other hand, a
near-threshold resonance (with $\Re(k)>- \Im(k)$) leads to a
cross section which grows with energy at $E=0$. If the enhancement
$f_0$ is further reduced, the S-matrix pole will recede from the
real energy axis at the same time losing impact on the scattering
cross section. The S-matrix pole trajectory is even more
complicated when presented in the energy manifold, see
Fig.~\ref{fig:trje}. At the critical point $E=0$ the physical pole
of the S-matrix is reflected backward while being projected into
the next Riemann sheet. At the point where physical and unphysical
poles collide, their trajectories turn by $\pi/2$ acquiring
 imaginary energy parts, whereas the real energy parts once
again start to increase. Obviously, such a non-analytic behavior is
highly non-linear and cannot be approximated by a simple
polynomial as was naively tried in \cite{PhysRevLett.118.232501}.

Returning back to the $4n$ system, we would like to stress
again the qualitative difference between  the $4n$ $J^\Pi = 0^+$ S-matrix pole
trajectories when bound dineutrons are produced from the ones when
dineutrons are kept unbound. In the first case the trajectory is
dominated by the two-dineutron threshold and thus inherits features
common to a two-body $S$-state one, i.e., the $4n$ bound state turns
into a two-dineutron virtual state. On the contrary, if no bound
dineutron states are present, at the critical point the $4n$ bound
state evolves directly into a narrow resonance -- this feature is
determined by the presence of the repulsive $1/r^2$ term in the
effective $4n$ hyperradial potential. Notably, such a
behavior is also exhibited by three-body Efimov states. On the right side of
the unitary limit (positive two-body scattering lengths) bound
Efimov states evolve from the virtual atom-dimer states. On the
contrary, on the left-hand side of the unitary limit (negative
two-body scattering lengths) bound Efimov states appear from
three-atom resonant states\cite{NaidonEndo}.

\section{Conclusions \label{s3}}

Using rigorous treatment of the four-particle continuum as
given by FY and AGS equations,
we investigated the $4n$ system with artificially enhanced interaction.
In contrast to our previous studies
\cite{PhysRevC.93.044004,lazauskas:4n,deltuva:18b} with no
bound dineutrons, here we enhanced also the two-neutron ${}^1S_0$ potential,
 aiming to elucidate the $0^+$ tetraneutron resonance behavior in the presence
of bound dineutrons, naturally appearing in the ${}^1S_0$ partial
wave once the factor $f_0$ exceeds roughly 1.1. As even larger
$f_0$ has been used in previous works suggesting the existence of
a near-threshold \cite{PhysRevLett.118.232501} or broad
\cite{PhysRevLett.119.032501} $0^+$ tetraneutron resonance, the
present study was aiming to resolve the  disagreement between
Refs.~\cite{PhysRevLett.118.232501,PhysRevLett.119.032501} and the
 $f_0 = 1$ approach \cite{PhysRevC.93.044004,lazauskas:4n,deltuva:18b}
contesting the presence of an observable $0^+$ tetraneutron
resonance.

Following the approach of
 Refs.~\cite{PhysRevLett.118.232501,PhysRevLett.119.032501}
with $f_0 = f_L =f$ we found that a significant enhancement
$f \approx 2.4$ to 2.7 is needed to support a truly bound tetraneutron
below the two-dineutron threshold. However, when $f$ is reduced the
$4n$ bound state evolves not into a resonance but into a virtual state.
This is convincingly demonstrated by the energy dependence of the
two-dineutron phase shifts. Thus, the negative energy $4n$ states,
considered in  Refs.~\cite{PhysRevLett.118.232501,PhysRevLett.119.032501}
to be zero-width bound states evolving into resonances at
positive energy, are neither bound states nor
narrow resonances.              
From the point of view of rigorous four-particle theory the
$E_{4n}<0$ states above the two-dineutron threshold are either
two-dineutron or dineutron plus two-neutron continuum states
without resonant character. With decreasing $f$ the scattering
observables approach  the universal zero-range behavior of the
four-fermion system with $E_d =0$. Note that on the way to the
physical potential strength the unitary limit has to be crossed.

We have also shown that enhancing the higher-wave potential
considerably stronger as compared to $S$ wave may lead to a
resonant $4n$ behavior, as demonstrated by the example of the
two-dineutron phase shift and cross section. Anyhow, well before reaching the $f_0 =
f_L$ limit, that resonance moves far away from the scattering
region, possibly below the two-dineutron threshold in the
unphysical sheet of the complex energy plane, thereby becoming
experimentally unobservable. This is consistent with the previous
studies \cite{PhysRevC.93.044004,lazauskas:4n,deltuva:18b}.
Another remarkable feature of this enhancement scheme is the
evolution of the bound state into a resonance via  the virtual
state appearing in a very narrow transition regime. A simple
two-body example exhibiting the same type of evolution was
presented. In both cases the essential feature is the presence of
a large negative effective range parameter.

\begin{acknowledgments}
The authors acknowledge discussions with J. Carbonell. A.D.
acknowledges support  by the Alexander von Humboldt Foundation
under Grant No. LTU-1185721-HFST-E. We were granted access to the
HPC resources of TGCC/IDRIS under the allocation 2018-A0030506006
made by GENCI (Grand Equipement National de Calcul Intensif).
\end{acknowledgments}


\end{document}